\documentclass[aps,pra,onecolumn]{revtex4-2}
\usepackage{amsmath}
\usepackage{amsfonts}
\usepackage{mathtools}  
\usepackage{color}
\definecolor{lavender}{rgb}{0.71, 0.49, 0.86}

\begin{document}

\title{Analytical connection between exact and approximate solutions of the periodically-driven two-level system starting from the Heun equation}

\author{Pietro Marco Follia}
\email{pietro.follia@unimi.it}
\affiliation{Dipartimento di Fisica ``Aldo Pontremoli", Università degli Studi di Milano, via Celoria 16, 20133 Milan, Italy}

\affiliation{Istituto Nazionale di Fisica Nucleare, Sezione di Milano, via Celoria 16, 20133 Milan, Italy}

\author{Bassano Vacchini}
\email{bassano.vacchini@mi.infn.it}
\affiliation{Dipartimento di Fisica ``Aldo Pontremoli", Università degli Studi di Milano, via Celoria 16, 20133 Milan, Italy}
\affiliation{Istituto Nazionale di Fisica Nucleare, Sezione di Milano, via Celoria 16, 20133 Milan, Italy}

\begin{abstract}
We investigate and establish an analytic connection between the exact solutions describing the dynamics of a two-level system driven by periodic external fields, focusing on the cases of linear driving and the so-called rotating-wave approximation, or circular driving. In both cases, the exact solutions can be obtained by mapping the Schrödinger equation onto Heun equations: the confluent Heun equation for linear driving and the Heun equation for the rotating-wave case. In particular, we demonstrate a direct analytic connection between the exact solutions for linear driving and those for the rotating-wave case. This result is obtained by analyzing local solutions expressed in terms of hypergeometric functions, which, in the case of the confluent Heun equation, can be derived by considering path-multiplicative Floquet solutions involving a bilateral series. This series leads to two continued-fraction expansions that can be perturbatively solved by imposing a suitable consistency condition.
The connection between the linear-driving and rotating-wave solutions is established through a perturbative procedure that allows us to recover not only the rotating-wave approximation itself, but also the correct Stark and Bloch–Siegert shifts, as well as the so-called high-frequency approximation.
\end{abstract}

\maketitle
\section{Introduction}
The dynamics of a two-level system (TLS) interacting with a periodic external driving field constitutes one of the most fundamental paradigms in quantum mechanics. Originally developed to understand magnetic resonance and light-matter interactions in quantum optics~\cite{Rabi1936a,Bloch1946a,Wangsness1953a,Shirley1965a,Allen1976}, this model has gained extraordinary prominence in recent years. Nowadays, it stands as the cornerstone of modern quantum technologies, where the driven TLS serves as the quintessential model for a qubit. Achieving precise analytical control over the evolution of such systems is crucial for high-fidelity quantum control, quantum computing architectures, and precision quantum metrology.

From a mathematical standpoint, the exact analytical description of these driven quantum systems has recently been connected to the realm of Heun differential equations. Just as the hypergeometric equation and its confluent forms constituted the mathematical backbone of theoretical physics throughout the twentieth century, the Heun equation is emerging as the relevant differential equation for many theoretical problems  of the twenty-first century. The Heun equation can be seen as a direct extension of the hypergeometric equation, generalizing it by admitting one additional singularity.  Over the last two decades, Heun equations have increasingly appeared in a vast and diverse array of cutting-edge physical contexts, including general relativity, conformal quantum field theory, astrophysics, solid-state physics, and fluid dynamics~\cite{Ishkhanyan2014a,Shahverdyan2015a,Staicova2016a,Hortacsu2018a}.

In the specific context of the periodically driven TLS, the
time-dependent Schrödinger equation can be exactly mapped onto
equations belonging to the Heun family. Specifically, it was first
shown in~\cite{Ma2007a,Xie2010a} and recently elaborated in~\cite{Schmidt2021a}  that this is the case for a linear driving field,
recovering a confluent Heun equation (CHE). 
The relevance of the  Heun equation (HE), also called general Heun equation, in connection with TLS in the presence of periodic potentials was later considered in~\cite{Xie2018a}. 
We here show that the HE, for a suitable choice of parameters, is directly associated to the Hamiltonian of a TLS for the case of driving in the rotating-wave approximation (RWA), also called circular driving since it describes the effect of a circularly polarized field.
Furthermore, a direct analytic connection bridging the linear-driving solutions with the rotating-wave solutions, to the best of our knowledge, is still missing. In this work, we close this gap. The core of our analysis  relies on the expression of local solutions in terms of hypergeometric
functions. In the case of the CHE, we show that these
representations can be derived by considering path-multiplicative
Floquet solutions involving a bilateral series. This series naturally
leads to two continued-fraction expansions, which are then solved by
imposing a suitable \textit{consistency condition}. 
We then consider different limiting situations, depending on the relationship between the driving strength, the system frequency, and the driving frequency. In the weak-driving approximation, we recover the correct Stark shift and Bloch-Siegert shift, as well as the RWA solution when the driving is resonant with respect to the system. Furthermore, the so-called high-frequency approximation, which leads to a frequency correction due to multiplication by a zeroth-order Bessel function, is also retrieved.

The paper is organized as follows. In Sec.~\ref{sec:General_on_Heun}, we introduce the essential mathematical properties of HE and CHE, detailing the relevant classes of solutions for the problems of interest. In Sec.~\ref{sec:exact}, we present the exact analytical solutions for linear and circular driving in terms of the Heun function. In Sec.~\ref{sec:analytical_connection}, we state our main results, providing the analytical connection between the exact solution for linear driving and different physical limits. A conceptual scheme summarizing the main results and connections discussed in this article is presented in Fig.~\ref{fig:scheme}. Finally, we discuss our results in Sec.~\ref{sec:ceo}. 

\section{Heun differential equations}
\label{sec:General_on_Heun}
The Heun equation, named after the German mathematician Karl Heun, is the second-order Fuchsian differential equation with 4 singularities. It can be written in the \textit{canonical form}~\cite{Ronveaux1995, Slavyanov2000a} as
\begin{equation}
\label{eq:HE}
    y''+ \left(\frac{\gamma}{z}+\frac{\delta}{z-1}+\frac{\epsilon}{z-a}\right)y' + \frac{\alpha\beta z-q}{z(z-1)(z-a)}y = 0,
\end{equation}
with poles at  $0,1,a,\infty$ and with the condition $\epsilon =\alpha +\beta-\gamma -\delta +1$ that grants regularity at infinity. The prime denotes the derivative with respect to $z$.
Heun differential equations are of great mathematical interest, since they encompass a wide variety of relevant special functions as their solutions and as solutions of their various confluent versions. Of particular interest for this manuscript is the equation that can be derived by a confluence process. The considered process of confluence is the symmetrization of two finite singularities: one singularity is moved at infinity making it irregular and leaving the differential equation with 3 singularities. Starting from Eq.~\eqref{eq:HE} it can be realized by $a \to +\infty$ with a proper rescaling of the other parameters and the resulting CHE can be written in the \textit{non-symmetrical canonical from}~\cite{Ronveaux1995,Slavyanov2000a}:
\begin{equation}
\label{eq:CHE}
    y'' +\left(\frac{\gamma}{z}+\frac{\delta}{z-1}+4p\right)y'+\frac{4 \alpha p z-\sigma}{z(z-1)}y = 0.
\end{equation}
Usually, $q$ and $\sigma$ are called \textit{accessory parameters}.
Additional confluence procedures give rise to other members of the Heun family, such as the double-confluent and bi-confluent equations. Furthermore, the hypergeometric equation can be obtained as special case of the HE by the confluence of the singularities in $a$ and $1$ and by setting $q = \alpha\beta$.
For recent work on these equations in the physical context we refer the interested reader to~\cite{Ishkhanyan2014a,Shahverdyan2015a,Staicova2016a,Hortacsu2018a}.

Although the characterization of the solutions of the Heun equation and its confluent forms remains largely incomplete, four main types of solutions are commonly distinguished, and three of them are presented below because of their relevance in our treatment.

At each regular singular point there exist two Frobenius solutions, which are convergent in a disk whose radius extends up to (but excluding) the nearest singularity. For the HE these are eight solutions and they can be identified by their characteristic exponent $r$, which determines the behaviour of the solution near the regular singularity. The characteristic exponents are $\{0, 1-\gamma\},\{0, 1-\delta\}, \{0, 1-\epsilon\}$ and $\{\alpha,\beta\}$ respectively at $0,1,a$ and $\infty$. The Frobenius solutions associated with a singularity at $z^*$ will be denoted by $H\ell^{(r)}_{\{z^*\}}(\boldsymbol{\mu},a;z)$, where ``$\ell$'' stands for local and with $\boldsymbol{\mu}=(\alpha,\beta,\gamma, \delta,\epsilon,q)$. These solutions of the HE admit a power-series expansion of the form
\begin{equation}
    H\ell^{(r)}_{\{z^*\}}(\boldsymbol{\mu},a ;z) = (z-z^{*})^r\sum_{n=0}^{\infty} b^{(r)}_n (z - z^*)^n,
\end{equation}
where the coefficients $b^{(r)}_n$ are determined by a three-term recurrence relation~\cite{Ronveaux1995,Slavyanov2000a}. In the case of the CHE the situation is more complex since the singularity at $\infty$ is an irregular point. It is still possible to define local solutions at $z^{*} = \{0,1\}$ with the characteristic exponents $\{0,1-\gamma\}$ and $\{0,1-\delta\}$ respectively. Following the same notation as for the HE they will be denoted by $H\ell^{(r)}_{\{z^*\}}(\boldsymbol{\eta};z)$ with $\boldsymbol{\eta} = (\alpha,\gamma, \delta,p,\sigma) $. The solutions of the CHE also admit a power-series expansion
\begin{equation}
\label{eq:power_exp_conf}
    H\ell^{(r)}_{\{z^*\}}(\boldsymbol{\eta};z) = (z-z^*)^{r}\sum_{n=0}^{\infty} d^{(r)}_n (z - z^*)^n,
\end{equation}
where the coefficients $d^{(r)}_n$ are still determined by a three-term recurrence relation~\cite{Ronveaux1995,Slavyanov2000a}.
Equivalently, the local solutions can be expanded in series of hypergeometric functions~\cite{Ronveaux1995,Slavyanov2000a,Schmidt1979a}, providing an alternative representation, which has traditionally received less attention. The expansion in terms of hypergeometric functions is at the core of this work and will be discussed in more detail in Sec.~\ref{sec:analytical_connection}.

If a Frobenius solution associated with a singularity $z^*$ can also be analytically continued to another singularity $w^*$, without developing singular behaviour in the intermediate domain, it is referred to as a Heun function, denoted by $Hf^{(r)}_{\{z^*,w^*\}}(\boldsymbol{\mu},a;z)$ and $Hf^{(r)}_{\{z^*,w^*\}}(\boldsymbol{\eta};z)$ respectively for HE and CHE, where ``$f$'' stands for function. This typically occurs only for specific (quantized) values of the accessory parameters.

Finally, the last class of solutions, which is central to this work, consists of path-multiplicative Floquet (PMF) solutions, also simply called path-multiplicative solutions. Given a closed path encircling exactly two singularities, such solutions, when analytically continued along the path, return to their initial value multiplied by a constant factor.
These solutions, that following Ronveaux~\cite{Ronveaux1995} will be simply denoted by $w(z)$, have been treated only in a few works, most importantly in Schmidt~\cite{Schmidt1979a}. While for Frobenius solutions of the HE it is straightforward to derive hypergeometric expansions, as will be clarified later in Sec.~\ref{sec:analytical_connection}, the existence of PMF solutions is necessary in order to obtain analogous series representations for local solutions of the CHE. 

\section{Exact analytical solutions for driven two-level systems in terms of Heun functions}
\label{sec:exact}
Recently, it has been shown that the Schr\"odinger problem of some periodically driven two-level system Hamiltonians can be mapped to Heun differential equations and some of its confluent forms~\cite{Ma2007a,Xie2010a,Xie2018a,Ishkhanyan2014a,Shahverdyan2015a}. We outline the general idea to realize this mapping. Consider the generic Hamiltonian
\begin{equation}
\label{eq:general_hamil}
    H(t) = \boldsymbol{f}(t)\cdot \boldsymbol{\sigma},
\end{equation}
where $\boldsymbol{f}(t) = \{f(t)_i\}_{i=x,y,z}$ is a vector of real-valued periodic functions with the same period $T = {2\pi}/{\omega}$,
\begin{equation}
\label{eq:periodo}
    \boldsymbol{f}(t+T)= \boldsymbol{f}(t)
\end{equation}
and where $\boldsymbol{\sigma} = \{\sigma_i\}_{i ={x,y,z}}$ are the Pauli matrices. The Schr\"odinger problem in the basis $\{|e\rangle, |g\rangle\}$ of $\sigma_z$'s eigenstates reads 
\begin{eqnarray}
\label{eq:shro_accop}
        &&i\frac{d}{dt}\psi_e(t) = f_{-}(t)\psi_g(t) + f_z(t)\psi_e(t) \nonumber \\
        &&i\frac{d}{dt}\psi_g(t) = f_{+}(t)\psi_e(t) -f_z(t)\psi_g(t).
\end{eqnarray}
For convenience, $f_{\pm}(t) = f_x(t)\pm if_y(t)$ were defined according to standard usage. It is important to note that, due to the self-adjointness of the Hamiltonian, the second differential equation in~\eqref{eq:shro_accop} follows directly from the first by complex conjugation together with the transformations $\psi_e(t)\to \psi_g(t)$ and $\psi_g(t)\to -\psi_e(t)$.
Eq.~\eqref{eq:shro_accop} constitutes a system of coupled first-order differential equations which can be rewritten as two autonomous second-order differential equations by taking the time derivative and by removing $\psi_g(t)$ and $\psi_e(t)$ respectively in the first and second equation by substitution. Changing variable to $\tau = \omega t$ this procedure results in the second-order differential equation
\begin{eqnarray}
\label{eq:shro_second}
    &&\ddot\psi_e(\tau) -\frac{\dot{f}_{-}(\tau)}{f_{-}(\tau)}\dot\psi_e(\tau) +\left[\frac{f_z(\tau)^2}{\omega^2}+\frac{|f_{-}(\tau)|^2}{\omega^2}+i\frac{\dot f_z(\tau)}{\omega}-i\frac{f_z(\tau)}{\omega}\frac{\dot f_{-}(\tau)}{f_{-}(\tau)}\right]\psi_e(\tau) = 0,
\end{eqnarray}
where the dot represents the derivative with respect to $\tau$.
The expression is well defined as long as ${\dot f_{-}(\tau)}/{f_{-}(\tau)}$ is well defined for all $\tau$. Once the differential equation for $\psi_e(\tau)$ is specified, the corresponding equation for $\psi_g(\tau)$ follows immediately by complex conjugation. Hence, without loss of generality, the analysis can be restricted only to the equation for $\psi_e(\tau)$.

The basic idea is to look for a transformation of variables $z = z(\tau)$ and an assignment
\begin{equation}
\label{eq:assignement}
    \psi_e(\tau) = \phi(z(\tau))\zeta(z(\tau))
\end{equation}
such that
that Eq.~\eqref{eq:shro_second} can be reduced to a known equation for $\zeta(z)$ of the form
\begin{equation}
\label{eq:general_second}
    \zeta(z)'' + P(z)\zeta(z)' + Q(z)\zeta(z) = 0.
\end{equation}
This allows to write the solutions of~\eqref{eq:shro_accop} in terms of known solutions of Eq.~\eqref{eq:general_second}.
In the scope of this manuscript are of particular interest the cases in which Eq.~\eqref{eq:general_second} corresponds to a HE or a CHE of the form~\eqref{eq:HE} or~\eqref{eq:CHE} respectively.
\subsection{Confluent Heun equation for linear driving}
\label{sec:che_ld}
It was first shown by  Ma and Xie~\cite{Ma2007a,Xie2010a} that the second-order differential equation for the linearly driven TLS can be mapped to a CHE. Consider the Hamiltonian
\begin{equation}
\label{eq:lin_TLS}
    \tilde{H}^{\text{\tiny L}}(t) = \frac{\omega_0}{2}\sigma_z + f\sin(\omega t)\sigma_x,
\end{equation}
where $L$ stands for linear. By performing the  unitary time-independent transformation 
\begin{equation}
\label{eq:unitaria}
    U=\exp{\left(i \frac{\pi}{3} \boldsymbol{n}\cdot \boldsymbol{\sigma}\right)}\qquad \mathrm{with} \qquad \boldsymbol{n} =\frac{1}{\sqrt{3}}(1,-1,1)
\end{equation}
it  becomes
\begin{equation}
\label{eq:linear_rota}
    {H}^{\text{\tiny L}}(t)= {U}^\dagger\tilde{H}^{\text{\tiny L}}(t){U} = f\sin(\omega t)\sigma_z-\frac{\omega_0}{2}\sigma_y,
\end{equation}
which provides the most suitable expression for the subsequent calculations.
Evidently, Eq.~\eqref{eq:linear_rota} is just a particular case of the Hamiltonian~\eqref{eq:general_hamil} with 
\begin{equation}
    \boldsymbol{f}^{\text{\tiny L}}(t) = \left(0, -\frac{\omega_0}{2}, f\sin(\omega t)\right),
\end{equation}
and its Schr\"odinger problem for $\psi_e(t)$ can be re-written as
\begin{equation}
    \ddot\psi_e(\tau) + \left[i\frac{f}{\omega}\cos(\tau)+\frac{f^2}{\omega^2}\sin^2(\tau)+\frac{\omega_0^2}{4\omega^2}\right]\psi_e(\tau) = 0.
\end{equation}
%
Finally, it is possible to define the variable transformation
\begin{eqnarray}\label{eq:defz}
    z(\tau): [0, \pi] &\longrightarrow& [0,1]  \nonumber\\ 
    \tau &\longmapsto& \sin^2\left(\frac{\tau}{2}\right),
\end{eqnarray}
and use the assignment Eq.~\eqref{eq:assignement} with
\begin{equation}
\label{eq:transform_var_lin}
   \phi^{\text{\tiny L}}(z) = \exp{\left(i\frac{2f}{\omega}z\right)}
\end{equation}
so that $\zeta^{\text{\tiny L}}(z)$ solves the CHE~\eqref{eq:CHE} with coefficients
\begin{equation}
\label{eq:params_confluent_heun}
    \alpha  = 1, \quad 
    \gamma = \delta = \frac{1}{2}, \quad 
    p = i\frac{f}{\omega} , \quad 
    \sigma =\frac{\omega_0^2}{4\omega^2}+ i\frac{2f}{\omega}.
\end{equation}
This explicit mapping was the first used to provide formal analytical solutions of the linearly driven TLS in terms of known solutions of the CHE~\cite{Ma2007a,Xie2010a}. The power-series expansion of the Frobenius solutions~\eqref{eq:power_exp_conf} was subsequently used in~\cite{Schmidt2021a} to study the Floquet theory of this system, focusing on the explicit expression of the quasienergies in terms of these solutions.

\subsection{Heun equation for driving in rotating-wave approximation}
\label{Sec:GHE_RWA}
\begin{figure}
 \centering
        \includegraphics[width=\textwidth]{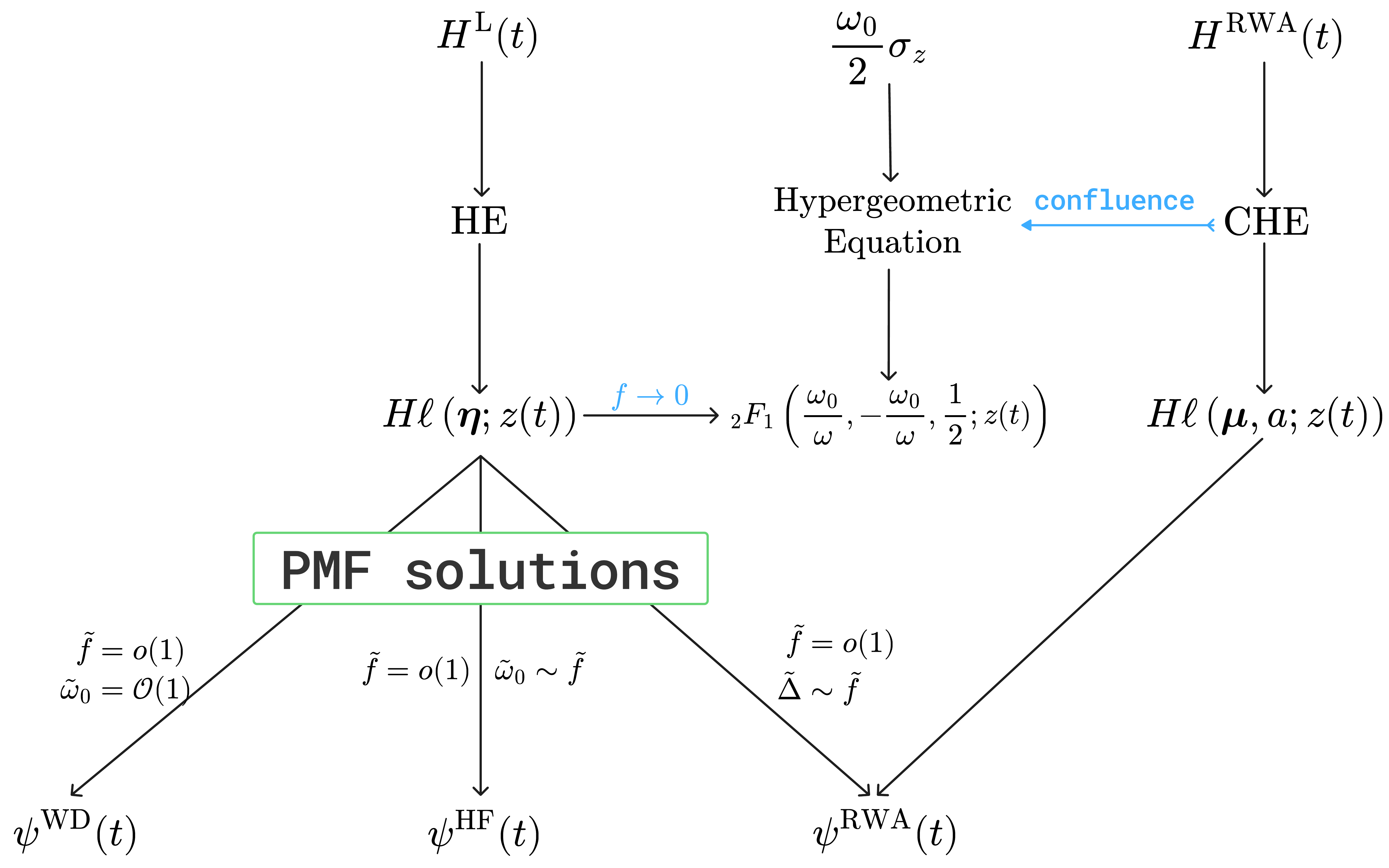}
 \caption{The figure provides a logical scheme of the connections between the different equations and their solutions and limits considered in the manuscript. Here, $\psi^{\text{\tiny WD}}(t)$ denotes the approximate solution in the weak-driving regime [see Eq.~\eqref{eq:PMF_weak_phase}], $\psi^{\text{\tiny HF}}(t)$ the high-frequency solution [see Eq.~\eqref{eq:High_freq_path}], and $\psi^{\text{\tiny RWA}}(t)$ the solution in the rotating-wave approximation [see Eq.~\eqref{eq:PMF_RWA_phase}]. According to Eqs.~\eqref{eq:ftilda}, \eqref{eq:omegazerotilda} and~\eqref{eq:rwalimit} we have $\tilde{f} ={f}/\omega$, $\tilde{\omega_0} ={\omega_0}/{2\omega}$ and $\tilde{\Delta} ={\Delta}/\omega$ respectively, that define the relevant limiting situations.}
\label{fig:scheme}
\end{figure}

The same procedure outlined above can be used to map the Schr\"odinger problem for the TLS with RWA driving to a HE, that corresponds to driving with a circularly polarized field or simply circular driving~\cite{Schmidt2018a}. Consider the RWA Hamiltonian
\begin{equation}
    \label{eq:RWA}
    \tilde{H}^{\text{\tiny RWA}}(t) = \frac{\omega_0}{2}\sigma_z +i\frac{f}{2}\left[\exp(-i\omega t)\sigma_+-\exp(i\omega t)\sigma_-\right],
\end{equation}
which by applying again the  unitary transformation~\eqref{eq:unitaria} can be written as
\begin{equation}
    H^{\text{\tiny RWA}}(t) = {U}^\dagger\tilde{H}^{\text{\tiny RWA}}(t){U}= \frac{f}{2}[\sin(\omega t)\sigma_z+\cos(\omega t)\sigma_x] -\frac{\omega_0}{2}\sigma_y.
\end{equation}
This is a particular case of the Hamiltonian~\eqref{eq:general_hamil} with
\begin{equation}
    \boldsymbol{f}^{\text{\tiny RWA}}(t) = \left(\frac{f}{2}\cos(\omega t),  -\frac{\omega_0}{2},\frac{f}{2}\sin(\omega t)\right).
\end{equation}
The first-order differential equation of the Schr\"odinger problem can be mapped to the second-order differential equation
\begin{equation}
    \ddot\psi_e^{{\text{\tiny RWA}}}(\tau) + \frac{\sin(\tau)}{\cos(\tau)+i({\omega_0}/{f})}\dot\psi_e^{{\text{\tiny RWA}}}(\tau) + \left[\frac{f^2+\omega_0^2}{4\omega^2}-\frac{\omega_0}{2\omega}\frac{\cos(\tau)-i({f}/{\omega_0})}{\cos(\tau)+i({\omega_0}/{f})} \right]\psi_e^{{\text{\tiny RWA}}}(\tau) = 0.
\end{equation}
Finally, it is possible to consider again the transformation~\eqref{eq:defz}  and simply take
\begin{equation}
\label{eq:defcircular}
   \phi^{{\text{\tiny RWA}}}(z) = 1
\end{equation}
in Eq.~\eqref{eq:assignement}, so that $\psi^{\text{\tiny RWA}}_e(\tau) =  \zeta^{\text{\tiny RWA}}(z(\tau))$.
The function $\zeta^{\text{\tiny RWA}}(z)$ then satisfies a HE of the form Eq.~\eqref{eq:HE} with coefficients that, upon defining
\begin{equation}
\label{eq:detuning}
    \Delta = \omega_0-\omega
\end{equation}
and
\begin{equation}
\label{eq:rabi}
    \Omega = \sqrt{f^2+\Delta^2},
\end{equation}
usually referred to as \textit{detuning parameter} and \textit{Rabi frequency} in the physical literature, are given by
\begin{eqnarray}
\label{eq:params_general_heun}
    &&\alpha = -\frac{1}{2} + \frac{\Omega}{2\omega}, \quad\beta =-\frac{1}{2} - \frac{\Omega}{2\omega},
  \nonumber \\ 
    &&\gamma = \delta = \frac{1}{2}, \quad
    \epsilon = -1, \nonumber \\ 
    &&a =\frac{1}{2} \left(1+i \frac{\omega_0}{f}\right),
\end{eqnarray}
together with
\begin{equation}
\label{eq:qqq}
    q =\left(\frac{\omega_0}{4\omega}-\frac{f(f+\omega_0)}{8\omega^2}\right) -i\left(\frac{\omega_0^3}{8\omega^2f}+\frac{f(2\omega+\omega_0)}{8\omega^2}\right).
\end{equation}
The solution of the Schr\"odinger problem with RWA Hamiltonian~\eqref{eq:RWA} are well-known and can be written as~\cite{Bloch1940a,Scully1997}
\begin{eqnarray}
\label{eq:solutions_rwa}
        &&\psi^{\text{\tiny RWA,I}}_e(t) = \frac{1}{2}\left[\left(1+\frac{if+\Delta}{\Omega}\right)\cos\left(\frac{\Omega}{2}t+\frac{\omega}{2}t\right)+\left(1-\frac{if+\Delta}{\Omega}\right)\cos\left(\frac{\Omega}{2}t-\frac{\omega}{2}t\right)\right], \nonumber \\
        &&\psi^{\text{\tiny RWA,II}}_e(t) = \frac{1}{2}\left[\left(1+\frac{if+\Delta}{\Omega}\right)\sin\left(\frac{\Omega}{2}t+\frac{\omega}{2}t\right)-\left(1-\frac{if+\Delta}{\Omega}\right)\sin\left(\frac{\Omega}{2}t-\frac{\omega}{2}t\right)\right],
\end{eqnarray}
where $\psi^{\text{\tiny RWA,I}}_e(t)$ denotes the solution with initial condition $\psi^{\text{\tiny RWA}}_e(0)=1$, while $\psi^{\text{\tiny RWA,II}}_e(t)$ the solution with initial condition $\psi^{\text{\tiny RWA}}_e(0)=0$. 
Nevertheless, the connection between these solutions known from standard quantum mechanics, 
and the analytical solutions of the HE is particularly instructive. In Sect.~\ref{sec:analytical_connection} it will be shown that~\eqref{eq:solutions_rwa} arise from the truncation of the hypergeometric expansion of the Frobenius solutions occurring when the accessory parameter $q$ satisfies a quantization constraint.
It is worth noticing that, according to Eq.~\eqref{eq:params_general_heun}, the limit $f/\omega_0\to 0$ implements a confluence process. This process, however, results in the hypergeometric equation that describes the free evolution of the TLS, as depicted in Fig.~\ref{fig:scheme}, and is therefore of no special interest.
We further notice that the mapping presented here is a generalization of the one introduced in~\cite{Xie2018a}, where only the case in which $\omega_0 = 0$ was considered, without showing  the connection of such a HE to the Schr\"odinger problem with RWA Hamiltonian~\eqref{eq:RWA}, that relies on the rotation introduced in Eq.~\eqref{eq:unitaria}.
Moreover in the present treatment we use the most simple assignement Eq.~\eqref{eq:defcircular}, rather than  $\phi^{{\text{\tiny RWA}}}(z) =(z-a)^2$ as in~\cite{Xie2018a}. Indeed, this additional factor can be reabsorbed by a transformation of the HE.

\section{Relevant physical approximations of the linearly driven system}
\label{sec:analytical_connection}
The mapping between the linearly driven TLS and the CHE allows us to study this system via exact analytical solutions of the confluent Heun equation. Power series expansion of Frobenius solutions around the regular singular point at $z^{*}=0$ have already been considered in previous works~\cite{Xie2010a,Schmidt2021a}. Indeed, using the variable transformation introduced in~\eqref{eq:defz}, and~\eqref{eq:transform_var_lin} as well as the power-series expansion~\eqref{eq:power_exp_conf}, the solutions of the linearly driven TLS can be written as
\begin{eqnarray}
    \psi_e(\tau) &=& \phi^{\text{\tiny L}}(\tau)H\ell^{(r)}_{\{0\}}(\boldsymbol{\eta} ;z(\tau)) 
    \nonumber \\
    &=& \exp{\left[i\frac{2f}{\omega}\sin^2\left(\frac{\tau}{2}\right)\right]}\sum_{n = 0}^{+\infty}d^{(r)}_n \sin^{2(n+r)}\left(\frac{\tau}{2}\right),
\end{eqnarray}
where the Frobenius solution with characteristic exponent $r=0$ yields the solution with initial condition $\psi_e(0)=1$, while the local solution with $r=1-\gamma$, which for $\gamma = 1/2$  is $r = 1/2$, corresponds to $\psi_e(0)=0$.
However, as pointed out in~\cite{Xie2010a,Schmidt2021a}, although exact, this representation of the solutions is not practical and cannot be used to recover known limiting regimes of the linearly driven TLS, such as the weak off-resonant driving regime, the RWA and the high-frequency limit.
We here employ the expansion in terms of hypergeometric functions~\cite{Ronveaux1995,Schmidt1979a,Erdelyi_1944a,Erdelyi_1942a,Slavyanov2000a} to recover these limiting cases directly within the exact analytical framework. For the reader's convenience, a conceptual overview of the obtained connections is provided in Fig.~\ref{fig:scheme}.

\subsection{Reference expression of solutions in the rotating-wave approximation}
\label{sec:exp_RWA_sol}
As anticipated, the Frobenius solutions of the HE~\eqref{eq:HE}  can also be represented as series of hypergeometric functions, which in general converge in a region containing only one singularity.
In particular,  the local solution around $z^* = 0$ with characteristic exponent $r = 0$ can be expanded as follows~\cite{Ronveaux1995, Svartholm_1939a, Erdelyi_1942a, Erdelyi_1944a}
\begin{equation}
\label{eq:exp_RWA_hyper}
    H\ell^{(0)}_{\{0\}}(\boldsymbol{\mu},a;z)=  \sum_{n=0}^{+\infty} c_n(\boldsymbol{\mu}) \, {_2}F_1(-\alpha - n + \gamma+ \delta -1, \alpha + n, \gamma, z),
\end{equation}
where ${_2}F_1(A,B,C,z)$ is the Gauss hypergeometric function and the coefficients $c_n(\boldsymbol{\mu})$ are determined by a three-term recurrence relation which, following the notation of~\cite{Ronveaux1995}, reads
\begin{equation}
\label{eq:rec_rel_hyper_RWA}
    P_n^{*} c_{n-1} + S_n^{*} c_n + R_n^{*} c_{n+1} = 0,
\end{equation}
with coefficients recalled in Appendix~\ref{appendix:A}.
For the parameters corresponding to the HE associated to the RWA driving given in Eq.~\eqref{eq:params_general_heun}, thanks to the
identity~\cite{Ronveaux1995,Slavyanov2000a}
\begin{equation}
\label{eq:ID_Hyp_Cos}
{_2}F_1\left(A,-A,\frac{1}{2}, \sin^2(x)\right) = \cos(2Ax)
\end{equation}
we are left with
\begin{equation}
      H\ell^{(0)}_{\{0\}}(\boldsymbol{\mu},a;z(\tau)) = \sum_{n= 0}^{+\infty} c_n(\boldsymbol{\mu})\cos((\alpha+n)\tau).
\end{equation}
Interestingly, the remaining accessory parameter $q$ appearing in the HE for the RWA can be obtained by asking that the series of hypergeometric functions appearing in Eq.~\eqref{eq:exp_RWA_hyper} is truncated at first order, thus defining a Heun function as discussed in Sec.~\ref{sec:General_on_Heun}. Indeed, in this case the relevant three-term  relations obtained from Eq.~\eqref{eq:rec_rel_hyper_RWA} are given by
\begin{eqnarray}
    && S^{*}_{0}c_{0}+R^{*}_{0}c_{1} = 0 \nonumber \\
    && P_1^{*}c_{0} + S^{*}_{1}c_{1} = 0 \nonumber \\
    && P_2^{*}c_{1} = 0.
\end{eqnarray}
Taking into account the definition of the coefficients recalled in Appendix~\ref{appendix:A} these conditions lead to the final constraint
\begin{equation}
    \frac{\alpha\beta}{4} = \left(\frac{\alpha}{2}+a\alpha^2+q\right)\left(\frac{\beta}{2}+a\beta^2+q\right),
\end{equation}
that for $\alpha,\beta$ and $a$ as in Eq.~\eqref{eq:params_general_heun} uniquely identifies the accessory parameter $q$ to be given by Eq.~\eqref{eq:qqq}.
The Frobenius solution is thus given by the Heun function
\begin{equation}
    Hf_{\{0,1\}}^{(0)}(\boldsymbol{\mu},a,z(t)) = c_0(\boldsymbol{\mu}) \cos\left( \frac{\Omega}{2}t-\frac{\omega}{2} t\right) + c_1(\boldsymbol{\mu}) \cos\left( \frac{\Omega}{2}t+\frac{\omega}{2} t\right),
\end{equation}
that using the explicit expressions of  $\alpha$ in Eq.~\eqref{eq:params_general_heun}, and recalling that the Ansatz Eq.~\eqref{eq:assignement} is verified with $\phi^{{\text{\tiny RWA}}}(z) = 1$ according to Eq.~\eqref{eq:defcircular}, recovers
the solution $\psi^{\text{\tiny RWA,I}}_e(t)$ of  Eq.~\eqref{eq:solutions_rwa}, corresponding to the system starting in the excited state.

\subsubsection{Alternative solution}
\label{sec:alternativa}
The other solution $\psi^{\text{\tiny RWA,II}}_e(t)$ can be obtained in a similar way starting from
the local solution with exponent $r=1-\gamma$. The latter is obtained according to the general theory~\cite{Ronveaux1995} by the relation
\begin{equation}
H\ell_{\{0\}}^{(1-\gamma)}(\boldsymbol{\mu},a;z)=z^{1-\gamma} H\ell^{(0)}_{\{0\}}(\boldsymbol{\mu}',a;z), 
\end{equation}
where the new coefficients are given by
\begin{eqnarray}
\label{eq:Coeff_RWA_ground_PRE}
    &&\alpha' = \alpha+1-\gamma,\quad \beta' = \beta+1-\gamma , \nonumber \\
    &&\gamma' = 2-\gamma, \quad \delta' = \delta, \quad  \epsilon' = \epsilon ,\nonumber  \\ 
     &&q' =q+ (a\delta +\epsilon)(1-\gamma) ,
\end{eqnarray}
leading to the identifications
\begin{eqnarray}
\label{eq:Coeff_RWA_ground}
    &&\alpha' =  \frac{\Omega}{2\omega},\quad \beta' =  -\frac{\Omega}{2\omega}, \nonumber \\
    &&\gamma' = \frac{3}{2}, \quad \delta' =  \frac{1}{2}, \quad \epsilon' = -1, \nonumber \\ 
     && q'  = \left(\frac{\omega_0}{4\omega}-\frac{f^2+\omega_0^2}{8\omega^2}-\frac38\right) -i\left(\frac{\omega_0^3}{8\omega^2f}+\frac{f(2\omega+\omega_0)}{8\omega^2}\right).
\end{eqnarray}
We can now consider an expansion of the local solution in hypergeometric functions in terms of the  transformed parameters
\begin{eqnarray}
    H\ell_{\{0\}}^{(1-\gamma)}(\boldsymbol{\mu},a;z) =z^{1-\gamma} \sum_{n=0}^{+\infty} c_n(\boldsymbol{\mu}') \, {_2}F_1(-\alpha' - n+ \gamma'+ \delta' -1, \alpha' + n , \gamma', z),
\end{eqnarray}
where the coefficients $c_n(\boldsymbol{\mu}')$ follow the same three-term recurrence relation  Eq.~\eqref{eq:rec_rel_hyper_RWA} with the transformed coefficients~\eqref{eq:Coeff_RWA_ground}.
Considering the latter expression for the coefficients and using  Eq.~\eqref{eq:defz} together with the identity~\cite{Ronveaux1995,Slavyanov2000a}
\begin{equation}
\label{eq:ID_Hyp_Sin}
_{2}F_1\left(A, 1-A, \frac{3}{2}, \sin^2(x)\right) = \frac{\sin((2A-1)x)}{(2A-1)\sin(x)}, 
\end{equation}
the Frobenius solution reads
\begin{eqnarray}
\label{eq:path_multiplicative_sin}
    H\ell_{\{0\}}^{(1/2)}(\boldsymbol{\mu},a;z(\tau)) =\sum_{n=0}^{+\infty} \frac{c_n(\boldsymbol{\mu}')}{2\alpha'+2n-1}\sin\left(\left(\alpha'+n-\frac{1}{2}\right)\tau\right).
\end{eqnarray}
It is immediate to verify that with the accessory parameter $q'$ given by Eq.~\eqref{eq:Coeff_RWA_ground} the recurrence for $c_n(\boldsymbol{\mu}')$ terminates as well at $n=1$, so that the Frobenius solution with $r = 1-\gamma$ is also a Heun function
\begin{eqnarray}
    Hf_{\{0,1\}}^{(1-\gamma)}(\boldsymbol{\mu},a;z(t))
    =  -\frac{\omega c_0(\boldsymbol{\mu}')}{\Omega+\omega}\sin\left( \frac{\Omega}{2}t-\frac{\omega}{2} t\right)-\frac{\omega c_1(\boldsymbol{\mu}')}{\Omega-\omega}\sin\left( \frac{\Omega}{2}t+\frac{\omega}{2} t\right).
\end{eqnarray}
Considering the explicit expression of $\alpha'$ given in Eq.~\eqref{eq:Coeff_RWA_ground} and using again the assignement Eq.~\eqref{eq:assignement} with $\phi^{\text{\tiny RWA}}(z) =1$ as in Eq.~\eqref{eq:defcircular} we recover the solution $\psi^{\text{\tiny RWA,II}}_e(t)$ of  Eq.~\eqref{eq:solutions_rwa}, corresponding to the system starting in the ground state.

It is important to notice that, while the mapping~\eqref{eq:defz} is bijective only in half of the period, the solutions obtained by expanding in terms of hypergeometric functions trivially extend to the entire time domain, differently from what happens when expressing the solution in power series.
The fact that the expansion in terms of hypergeometric functions is the most natural one when dealing with the TLS in RWA driving suggests that this choice can be the most convenient also when dealing with the linearly driven TLS.
As anticipated in Sec.~\ref{sec:General_on_Heun}, defining an analogous expansion for the local solutions of the CHE is not as straightforward, and the use of PMF solutions is necessary.

\subsection{Path-multiplicative Floquet solutions}
\label{sec:exp_PMF_sol}
Consider a smooth closed path $\Gamma$ in $\mathbb{C}$ encircling two, and only two, singularities of HE or CHE
\begin{eqnarray}
    \Gamma: [0,1] \longrightarrow \mathbb{C} \quad \text{s.t.} \quad \Gamma(0) = \Gamma(1) , \quad \dot\Gamma(t)\neq 0 ~\forall ~t\in[0,1].
\end{eqnarray}
A path-multiplicative Floquet solution with respect to $\Gamma$ is a solution which, when continued analytically around that path returns to its starting point multiplied by a constant~\cite{Ronveaux1995,Slavyanov2000a}
\begin{equation}
\label{eq:def_path_multi}
    w(\Gamma(1)) = e^{2\pi \nu i} w(\Gamma(0)), 
\end{equation}
where $\nu$ is called \textit{path exponent}. Given the path it can be proven that such solutions always exist. Furthermore, it can be proven that they are equivalent for all homotopic paths, so that to a path $\Gamma$ around only two singularities corresponds a unique set of PMF solutions.
Clearly, the value of $\nu$ is multiply-defined, since $\nu + 2\pi n$ with $n\in \mathbb{Z}$ still satisfies Eq.~\eqref{eq:def_path_multi}.
Consider the case of the CHE and a path encircling the two regular poles at $\{0,1\}$. Then there exist two linearly independent path-multiplicative solutions. Schmidt~\cite{Schmidt1979a} proposed and showed that it is possible to express these solutions, in an elliptic anulus with foci $0,1$ as an expansion in terms of hypergeometric functions, namely
\begin{equation}
    w(z) = \sum_{\nu\in \Xi} k_\nu z^{\nu} {_2}F_1(-\nu, 1-\gamma-\nu, 2-\gamma-\delta-2\nu, z^{-1} ),
\end{equation}
where $\Xi\subset \mathbb{C}$ is the set of all $\nu$ satisfying Eq.~\eqref{eq:def_path_multi}.
It is possible to show that as long as $2\nu -2+\gamma +\delta \notin \mathbb{Z}$, then $-\nu+1-\gamma-\delta$ is another path exponent, a condition typically verified in applications of physical interest.

According to a theorem proven by Schmidt~\cite{Schmidt1979a}, the existence of PMF solutions allows for the decomposition of the Frobenius solutions in a double sided expansion in terms of hypergeometric functions~\cite{Ronveaux1995}, in particular for the solutions in $z^*=0$ we have
\begin{equation}
\label{eq:exp_path_linear}
    H\ell^{(0)}_{\{0\}}(\boldsymbol{\eta};z) = \sum_{n\in \mathbb{Z}} g_n(\boldsymbol{\eta}) {_2}F_1(-n-\nu, n+\nu+\delta+\gamma-1, \gamma, z),
\end{equation} 
where the coefficients $g_n(\boldsymbol{\eta})$ satisfy a three terms recurrence relationship
\begin{equation}
\label{eq:recurrence_rel}
    A_ng_{n-1} + B_ng_n+ C_ng_{n+1} = 0,
\end{equation}
where again we follow the notation of~\cite{Ronveaux1995} for the coefficients, explicitly recalled in Appendix~\ref{appendix:A}.
Unfortunately, the path exponents $\nu$ do not have a known general expression in terms of  the parameters of the CHE~\eqref{eq:CHE}. In Sect.~\ref{sec:perturba} we will therefore introduce perturbative  methods to obtain their values. 
We further stress that, while both HE and CHE admit an expansion of the considered solutions  in terms of hypergeometric functions, the series~\eqref{eq:exp_RWA_hyper} for the expansion of the solution of the HE is single-sided, while the series~\eqref{eq:exp_path_linear} we just derived for the solution of the CHE is double-sided. The three-term recurrence relation determining the coefficients for the HE equation, namely Eq.~\eqref{eq:rec_rel_hyper_RWA}, must be satisfied for $n\in\mathbb{N}$, while the corresponding three-term recurrence for the CHE equation Eq.~\eqref{eq:recurrence_rel} must be satisfied for $n\in\mathbb{Z}$. 

\subsection{Perturbative expansion via continued fractions}
\label{sec:perturba}
The Frobenius solutions of the CHE associated with the TLS with linear driving are here considered. 
With the coefficients given in Eq.~\eqref{eq:params_confluent_heun} the expansion of the Frobenius solution~\eqref{eq:exp_path_linear} thanks to Eq.~\eqref{eq:ID_Hyp_Cos} becomes
\begin{equation}
\label{eq:path_multiplicative}
    H\ell^{(0)}_{\{0\}}(\boldsymbol{\eta};z(\tau)) = \sum_{n \in \mathbb{Z}} g_n(\boldsymbol{\eta}) \cos\left(\left(n+\nu\right)\tau\right),
\end{equation}
while the coefficients of the three-term recurrence relation according to Appendix~\ref{appendix:A} read
\begin{eqnarray}
\label{eq:rec_rel_ceoff_lin}
&& A_n = -C_n = -i\tilde{f}(n+\nu), \nonumber \\
&& B_n = (n+\nu)^2 - \tilde{\omega}_0^2,   
\end{eqnarray}
where the adimensional coefficients
\begin{equation}
\label{eq:ftilda}
    \tilde{f} =\frac{f}\omega 
\end{equation}
and 
\begin{equation}
\label{eq:omegazerotilda}
 \tilde{\omega}_0 = \frac{\omega_0}{2\omega}
\end{equation}
have been defined for the sake of convenience.
As follows also from Eqs.~\eqref{eq:path_multiplicative} and~\eqref{eq:rec_rel_ceoff_lin}, if $\nu$ is a path exponent then this also holds for $-\nu$. For this reason we will restrict our analysis to  the positive solutions only.
With the aim of expanding the local solutions to find known limiting cases it is useful to bring the three terms recurrence relation~\eqref{eq:recurrence_rel} in continued fraction form~\cite{Baber1935a,Leaver1985a}. Because the series is double-sided, it is necessary to introduce two continued fractions: one valid for $n \ge 0$ and one for $n \le 0$. This is achieved by defining
\begin{eqnarray}
\label{eq:ref_R_L}
    R_{n+1} = \frac{g_{n+1}}{g_n} \quad n \ge 0, \nonumber  \\
    L_{n-1} = \frac{g_{n-1}}{g_n} \quad n \le 0 .
\end{eqnarray}
By substituting these definitions into Eq.~\eqref{eq:recurrence_rel} we obtain
\begin{equation}
\label{eq:conti_fractionsR}
    R_{n} = \frac{{i\tilde{f}(\nu+n)}}{{(\nu+n)^2 - \tilde{\omega}_0^2} + {i\tilde{f}(\nu+n)}R_{n+1}} 
\end{equation}
and
\begin{equation}
\label{eq:conti_fractionsL}
    L_{n} = \frac{{i\tilde{f}(\nu+n)}}{-{(\nu+n)^2 + \tilde{\omega}_0^2} + {i\tilde{f}(\nu+n)}L_{n-1}}, 
\end{equation}
respectively.
These equations imply in particular that $\{R_n\}$  and $\{-L_n\}$ obey the same recurrence relation, so that in what follows it will be enough to determine the sequence $\{R_n\}$. We recall, however, that $R_n$ and $L_n$ are only defined for  positive and negative integers respectively.
Suppose that the exact solutions for $R_n$ and $L_n$ are available. In this case, the three-term recurrence relation is  satisfied  except at $n=0$, where the two solutions must be matched. 
Considering Eq.~\eqref{eq:recurrence_rel} for $n=0$ and dividing by $g_0$ we obtain
\begin{equation}
\label{eq:condizione}
    A_0 L_{-1} + B_0 + C_0 R_1 = 0.
\end{equation}
This condition expresses a constraint that we write in the compact form
\begin{equation}
\label{eq:consist_cond}
    F(\tilde{f}, \tilde{\omega}_0, \nu) = 0,
\end{equation}
and will be referred to hereafter as  \textit{consistency condition}. Since the path exponent is the only parameter that is not pre-determined, this condition becomes a constraint on the admissible values of $\nu$. It can thus be interpreted as a quantization condition arising from the two continued fractions of Eqs.~\eqref{eq:conti_fractionsR} and~\eqref{eq:conti_fractionsL}, together with the appropriate boundary conditions.

We will now use the continued-fraction representation of Eqs.~\eqref{eq:conti_fractionsR} and~\eqref{eq:conti_fractionsL} to construct approximate explicit expressions for $R_n$ and $L_n$ in different regimes of $\tilde{f}$ and $\tilde{\omega}_0$, further using the \textit{consistency condition} to obtain an approximate expression for the path exponent $\nu$, which in turn determines the solution.

\subsection{Weak-driving case}
\label{sec:weakd}
The regime in which $f/\omega \ll 1$ while $\omega \sim \omega_0$ is referred to as the weak-driving regime. These conditions can be expressed as $\tilde{f} = o(1)$ and $\tilde{\omega}_0 = \mathcal{O}(1)$. To consider the most general case, we assume that the path exponents and the continued fractions are functions of both $\tilde{f}$ and $\tilde{\omega}_0$. We can thus expand them in series as 
\begin{eqnarray}
\nu &=& \nu^{(0)} + \nu^{(1)} + \nu^{(2)}+ ...\, ,\nonumber \\
R_n &=& R_n^{(0)}+R_n^{(1)}+R_n^{(2)}+...\, ,\nonumber\\
L_n &=& L_n^{(0)}+L_n^{(1)}+L_n^{(2)}+...\, ,    
\end{eqnarray}
 where $\nu^{(i)},R_n^{(i)},L_n^{(i)} \sim \tilde{f}^i$. By substituting these expressions into the continued fraction~\eqref{eq:conti_fractionsR} we obtain
\begin{equation}
\label{eq:continued_explicit}
    R_n = i\tilde{f}(n+ \nu^{(0)}+\nu^{(1)}+...)\frac{1}{\left(n+\nu^{(0)}\right)^2-\tilde{\omega}_0^2+(n+\nu^{(0)})(2\nu^{(1)}+i\tilde{f} R_{n+1}^{(0)})+...},
\end{equation}
where the elements in the denominator have been grouped by their exponential behaviour with respect to $\tilde{f}^i$. 
Note that the connection between the coefficients in the continued fractions also holds perturbatively, so that we have $L_n^{(i)}=-R_{\lvert n\rvert}^{(i)}$. 
Introducing the functions
\begin{equation}
h_n(\nu^{(0)}, \tilde{\omega}_0)\coloneqq \left(n+\nu^{(0)}\right)^2-\tilde{\omega}_0^2,
\end{equation}
and assuming that $h_n(\nu^{(0)}, \tilde{\omega}_0) = \mathcal{O}(1)$ for all $n$,  the continued fraction expressions can be expanded as
\begin{eqnarray}
\label{eq:first_exp_weak}
    R_n &=& i\tilde{f}(n+ \nu^{(0)}+\nu^{(1)}+...)\left[\frac{1}{h_n(\nu^{(0)},\tilde{\omega}_0)}-\frac{(n+\nu^{(0)})(2\nu^{(1)}+i\tilde{f} R_{n+1}^{(0)})}{h_n(\nu^{(0)},\tilde{\omega}_0)^2}+...\right].
\end{eqnarray}
\subsubsection{Zeroth-order solution}
The zeroth-order terms are given by $L_{n}^{(0)}=- R_{|n|}^{(0)} =0 $, which implies  $g^{(0)}_n= 0$ for all $n\not = 0$. 
By substituting this result into the \textit{consistency condition}~\eqref{eq:consist_cond} we obtain according to Eqs.~\eqref{eq:rec_rel_ceoff_lin} and~\eqref{eq:condizione}  the result $B^{(0)}_0 = (\nu^{(0)})^2 - \tilde{\omega}_0^2 = 0$, implying for the expression of the two path exponents of order zero
\begin{equation}
\label{nuordero}
   \nu^{(0)}\omega = \pm \frac{\omega_0}{2}.
\end{equation}
In order to connect the zeroth-order Frobenius solution to the physical solution, we use the assignment~\eqref{eq:assignement}, 
that according to Eqs.~\eqref{eq:defz} and~\eqref{eq:transform_var_lin} involves
an additional $T$-periodic phase, namely
\begin{equation}
    \psi^{\text{I}}_e(t) = \exp{\left[ i\frac{2f}{\omega}\sin^2\left(\frac{\omega t}2\right)\right]}H\ell^{(0)}_{\{0\}}\left(\boldsymbol{\eta};\sin^2\left(\frac{\omega t}2\right)\right).
\end{equation}
In the weak-driving regime i.e., $\tilde{f} =  o(1)$, at zeroth-order the exponential can be replaced by unit and we obtain 
\begin{equation}
\label{eq:zeroth_order_sol_weak}
\psi^{\text{I}}_e(t) =  g^{(0)}_0\cos\left(\frac{\omega_0}{2}t\right),
\end{equation}
which is the well-known solution for the TLS without drive~\cite{Xie2010a}. 
\subsubsection{First-order solution}
To proceed at next order, we substitute  $R_n^{(0)} = 0$ in the expansion~\eqref{eq:first_exp_weak}, thus obtaining
\begin{equation}
\label{eq:continued_first_order_weak}
    R^{(1)}_n =\frac{i\tilde{f}(n+ \tilde{\omega}_0)}{h_n(\pm\tilde{\omega}_0,\tilde{\omega}_0)}.
\end{equation}
Inserting this expression along with the corresponding one for $L^{(1)}_n$  into the \textit{consistency condition}~\eqref{eq:consist_cond} gives the first-order approximation for the path exponents 
\begin{equation}
\nu^{(1)}\omega = \pm \frac{\omega_0}{2}\frac{f^2}{(\omega_0^2-\omega^2)}.
\end{equation}
Up to first-order term we thus obtain
the  expression
\begin{equation}
    \nu_{\pm}\omega = \pm(\nu^{(0)}+ \nu^{(1)})\omega + \textit{o}(\tilde{f}^3)= \pm  \left[ \frac{\omega_0}{2} + \frac{1}{4} \left( \frac{1}{\omega_0 - \omega} + \frac{1}{\omega_0 + \omega} \right) f^2 \right] + \textit{o}(\tilde{f}^3), 
\end{equation}
that we have written in such a way as to put into evidence the emergence of the well-known Stark and Bloch-Siegert shifts~\cite{Bloch1940a,Autler1955a}.
As anticipated, $g_0$ is the only coefficient contributing to zeroth-order. 
To proceed further, we have to consider the first-order approximations for $R^{(1)}_n$ and $L^{(1)}_n$  implied by Eq.~\eqref{eq:continued_first_order_weak}, together with the definition~\eqref{eq:ref_R_L}. We find
\begin{eqnarray}
    g_1^{(1)} = R^{(1)}_1\left(g^{(0)}_0+g_0^{(1)} \right)\sim \tilde{f},\nonumber \\ 
    g_{-1}^{(1)} = L^{(1)}_{-1}\left(g^{(0)}_0+g_0^{(1)} \right) \sim \tilde{f}.
\end{eqnarray}
By iteration, it follows that $g_{\pm n} \sim \tilde{f}^n$, so that only $g_0^{(0)}+g_0^{(1)}$ and $g_{\pm 1}^{(1)}$ need to be retained within the first-order approximation.
Starting from the local Frobenius solution at first-order, we obtain the physical solution again relying on Eq.~\eqref{eq:assignement} but keeping for the  exponential in~\eqref{eq:transform_var_lin} terms up to the first-order, thus obtaining 
\begin{equation}
\label{eq:PMF_weak_phase}
    \psi^{\text{I}}_e(t)  = a_0\cos(\nu_+ \omega t)+ a_1\cos( (\nu_+ +1) \omega t) +a_{-1}\cos( (\nu_+ -1) \omega t),
\end{equation}
where the coefficients are given by
\begin{eqnarray}
    &&a_0 = \left(1+ i\frac{f}{\omega} \right)g_0^{(0)}+g_0^{(1)}, \nonumber \\ 
    &&a_{\pm1} = -i\frac{f}{2\omega}g_0^{(0)}+g_{\pm1}^{(1)}.
\end{eqnarray}
It is important to note that the expression obtained for $\nu$ is valid only for $\omega \neq \omega_0$ . This limitation is not an artifact of the method, but rather reflects the breakdown of the condition $h_n(\nu^{(0)}, \tilde{\omega}_0) = \mathcal{O}(1)$ for all $n$. The situation in which this condition fails will hereafter be referred to as the \textit{resonant condition}.
\subsection{Resonant condition and rotating-wave approximation}
Considering  the detuning parameter introduced in Eq.~\eqref{eq:detuning} we can write
\begin{equation}
\label{eq:rwalimit}
\tilde{\omega}_0 = \frac{1}{2} + \frac{\tilde\Delta}{2},
\end{equation}
with ${\tilde\Delta}={\Delta}/{\omega}$. In the  perfectly resonant case, so that $\omega = \omega_0$, we obtain  $\tilde{\omega}_0 = \frac{1}{2}$, hence for the zeroth-order contribution  $\nu^{(0)} = \pm \frac{1}{2}$, implying 
$h_{\mp 1}(\nu^{(0)} , \tilde{\omega}_0)=0$.  Similarly, for the near-resonant case, in which $\tilde{\Delta}\sim \tilde{f}$, for the zeroth-order we obtain $h_{\mp 1}(\nu^{(0)} , \tilde{\omega}_0)\sim \tilde{f}$. In both cases, the condition $h_n(\nu^{(0)}, \tilde{\omega}_0)=  \mathcal{O}(1)$ for all $n$ breaks down, and the previously proposed expansion is no longer well defined for $L_{-1}$ and $R_1$, respectively when $\nu^{(0)} = 1/2$ and $\nu^{(0)} = -1/2$.
We are thus led to follow an alternative path.

For simplicity, we can  restrict the subsequent analysis to the case $\nu^{(0)} = \frac{1}{2}$ as the case $\nu^{(0)} = -\frac{1}{2}$ can be treated analogously. 
In this situation, a modified \textit{consistency condition} must be introduced, which does not require the definition of $L_{-1}$. 
To this end, consider the two consecutive three-term recurrence relations centred at $n=0$ and $n=-1$:
\begin{eqnarray}
\label{eq:condition_RWA}
    \begin{cases}
        A_0 g_{-1} + B_0 g_0 + C_0 g_1 = 0 \\
        A_{-1} g_{-2} + B_{-1} g_{-1} + C_{-1} g_0 = 0.
    \end{cases}
\end{eqnarray}
The expansion proposed in~\eqref{eq:first_exp_weak} is well defined for $L_{-2}$ and $R_1$, so that we can  write $g_1 = R_1 g_0$ and $g_{-2} = L_{-2} g_{-1}$. Substituting into Eq.~\eqref{eq:condition_RWA} yields
\begin{eqnarray}
    \begin{cases}
        A_0 g_{-1} + (B_0 + C_0 R_1) g_0 = 0 \\
        (A_{-1} L_{-2} + B_{-1}) g_{-1} + C_{-1} g_0 = 0
    \end{cases}
\end{eqnarray}
which form a system of two coupled equations for $g_{-1}$ and $g_0$. 
Non-trivial solutions
exist only if the determinant of the coefficients of the linear system is non-vanishing,
which leads to the condition
\begin{equation}
\label{eq:cons_cond_RWA}
   A_0 C_{-1} - (B_0 + C_0 R_1)(A_{-1} L_{-2} + B_{-1}) = 0.
\end{equation}
It is straightforward to verify that, whenever $L_{-1}$ is well defined, this condition is equivalent to the \textit{consistency condition} defined in~\eqref{eq:consist_cond}.
It is now possible to use this condition to derive the expansion of the path exponents in the weak-driving and near-resonant regime by substituting the expansions defined by Eq.~\eqref{eq:first_exp_weak} for $L_{-2}$ and $R_{1}$ into the \textit{consistency condition} defined in Eq.~\eqref{eq:cons_cond_RWA}. 
As expected, at zeroth-order we find 
\begin{equation}
    \nu^{(0)} \omega= \frac{\omega}{2}
\end{equation}
while at first-order in $\tilde{f}$, we obtain
\begin{equation}
    \nu^{(1)} \omega=  \frac{\Omega}{2}.
\end{equation}
By treating analogously the case $\nu^{(0)} = -\frac{1}{2}$, we find for the the path exponents up to first-order
\begin{equation}
    \nu_{\pm}\omega 
    = \pm \left(\frac{\omega}{2} + \frac{\Omega}{2}\right)+ \textit{o}(\tilde{f}^2).
\end{equation}
It is important to notice that in the resonant case, contrary to Sect.~\ref{sec:weakd}, $g_0$ is not the only term contributing to zeroth-order. Indeed, from the definitions~\eqref{eq:conti_fractionsR} and \eqref{eq:conti_fractionsL} it is immediate to verify that $L_{-1} = \mathcal{O}(1)$ for $\nu^{(0)} = 1/2$,  and $R_{1}= \mathcal{O}(1)$ for $\nu^{(0)}= -1/2$. This is a direct consequence of the condition $h_n(\nu^{(0)}, \tilde{\omega}_0) = \mathcal{O}(1)$ breaking down in the resonant case for $n=-1$ and  $n=1$ respectively. In this situation clearly
\begin{equation}
    g_{-1}^{(0)} = L_{-1}g_0^{(0)} =  \mathcal{O}(1),
\end{equation}
so that for the other coefficients it follows, by iteration, that $g_{n}\sim \tilde{f}^n$, $g_{-n}\sim \tilde{f}^{n-1}$.
To first order one only retains the term proportional to $g^{(0)}_0+g_{0}^{(1)}$, $g^{(0)}_{-1}+g^{(1)}_{-1}$, $g_{1}^{(1)}$ and $ g_{-2}^{(1)}$, thus obtaining, according to the same procedure used in  Sect.~\ref{sec:weakd}, the expression
\begin{equation}
\label{eq:PMF_RWA_phase}
    \psi^{\text{I}}_e(t)  = a_0\cos(\nu_+ \omega t)+ a_{-1}\cos( (\nu_+ -1) \omega t) +a_{1}\cos( (\nu_+ +1) \omega t) + a_{-2}\cos( (\nu_+ -2) \omega t),
\end{equation}
where we have defined the coefficients
\begin{eqnarray}
    &&a_0 = \left(1+i\frac{f}{2\omega} \right)g_0^{(0)}+ g_0^{(1)}-i\frac{f}{2\omega}g^{(0)}_{-1}, \nonumber \\ 
    &&a_{-1} = \left(1+i\frac{f}{2\omega} \right) g_{-1}^{(0)}+g_{-1}^{(1)}-i\frac{f}{2\omega}g^{(0)}_{0} , \nonumber\\ 
    &&a_{1} = -i\frac{f}{2\omega}g^{(0)}_{0} + g^{(1)}_{1}, \nonumber \\
    &&a_{-2} = -i\frac{f}{2\omega}g^{(0)}_{-1}+g^{(1)}_{-2},
\end{eqnarray}
so that we indeed recover the well-known rotating wave approximation~\cite{Rabi1936a,Shirley1965a}. It is important to observe that, as a consequence of the resonance, the coefficients of order $\mathcal{O}(1)$ are those associated to the frequencies $\nu_+\omega$ and $(\nu_+-1)\omega$, that thus provide the most important contribution to the dynamics.
As expected from the standard theory, the RWA solutions are a first-order approximation to the linearly driven TLS in the resonant condition under weak-driving i.e., $\tilde\Delta \sim \tilde{f}$, $\tilde{f}=o(1)$.
\subsection{High-frequency approximation}
Finally, the high-frequency limit can be derived by considering $\tilde{f} \sim \tilde{\omega}_0$, $ \tilde{f} =  o(1)$ and performing an expansion of the continued fraction in Eq.~\eqref{eq:continued_explicit}, which yields
\begin{equation}
    R_n = i\tilde{f}(n + \nu^{(0)} + \nu^{(1)} + \cdots)\left[\frac{1}{\left(n+\nu^{(0)}\right)^2} - \frac{2\nu^{(1)}+i\tilde{f} R^{(0)}_{n+1}}{\left(n+\nu^{(0)}\right)^3} + \cdots \right].
\end{equation}
At zeroth-order, one has $L^{(0)}_{n}=-R^{(0)}_{|n|}= 0$, which, upon substitution into the \textit{consistency condition}, yields $\nu^{(0)} = 0$. 
At first-order, we find 
\begin{equation}
    R^{(1)}_n  = \frac{i\tilde{f}}{n},
\end{equation}
leading thanks to the \textit{consistency condition} to $\nu^{(1)} = \pm \tilde{\omega}_0$. 
Proceeding to second-order, we obtain
\begin{equation}
    R^{(2)}_n = - \frac{i\tilde{f}\tilde{\omega}_0}{n^2},
\end{equation}
finally yielding for the path exponent up to second-order
\begin{equation}
\label{eq:High_freq_path}
    \nu_{\pm}\omega = \pm\frac{\omega_0}{2} \left(1- \frac{f^2}{\omega^2}\right)+\textit{o}(\tilde{f}^3).
\end{equation}

At this level of approximation, it is sufficient to retain only the first-order terms in the expansion of the PMF solutions obtaining the same formal expression as per the weak-driving case, see Eq.~\eqref{eq:PMF_weak_phase}.
This result is consistent with the standard high-frequency approximation, where the dynamics is governed by an effective frequency given by~\cite{Shirley1965a,Holthaus1992a}
\begin{equation}
    \omega_{\mathrm{eff}} = \frac{\omega_0}{2} J_0\left(\frac{2f}{\omega}\right),
\end{equation}
once we recall the relevant expansion for the Bessel function
\begin{equation}
    J_0\left(x\right) = 1 - \frac{x^2}{4} + o(x^4).
\end{equation}

\subsection{Alternative solutions}
\label{sec:alternative}
For all the cases detailed  above we can  proceed in a similar way as in Sec.~\ref{sec:alternativa} in order to determine the expression for the Frobenius solution with characteristic exponent $r =1-\gamma$. 
We start from 
\begin{equation}
     H\ell^{(1-\gamma)}_{\{0\}}(\boldsymbol{\eta};z(\tau)) = z^{1-\gamma}H\ell^{(0)}_{\{0\}}(\boldsymbol{\eta}' ;z(\tau)),   
\end{equation}
with transformed parameters given now by the relations
\begin{eqnarray}
\label{eq:coeff_linear_ground}
    &&\alpha' = \alpha +1-\gamma  , \nonumber\\
    &&\gamma' = 2-\gamma , \quad \delta' = \delta , \quad p' = p,  \nonumber\\ 
    &&\sigma' =\sigma+(\gamma-1) (\delta-4p).
\end{eqnarray}
Inserting these values in the expansion Eq.~\eqref{eq:exp_path_linear} and using Eq.~\eqref{eq:ID_Hyp_Sin} we come to
\begin{equation}
    H\ell^{(1/2)}_{\{0\}}(\boldsymbol{\eta};z(\tau)) = \sum_{n\in \mathbb{Z}} \frac{g_n(\boldsymbol{\eta}')}{2\nu'+2n+1} \sin\left(\left (\nu'+\frac{1}{2}+n\right)\tau\right),
\end{equation}
where the coefficients $g_n(\boldsymbol{\eta}')$ obey Eq.~\eqref{eq:recurrence_rel}.
This solution is still in the form of Eq.~\eqref{eq:path_multiplicative_sin}.
Defining $\nu = \nu'+1/2$ and $g_n(\boldsymbol{\eta})=g_n(\boldsymbol{\eta}')/(2\nu'+2n+1)$ we obtain for $g_n(\boldsymbol{\eta})$ the same recurrence relationship as in~\eqref{eq:rec_rel_ceoff_lin} so that 
\begin{equation}
    H\ell^{(1/2)}_{\{0\}}(\boldsymbol{\eta};z(\tau)) = \sum_{n\in \mathbb{Z}} g_n(\boldsymbol{\eta}) \sin\left(\left (\nu+n\right)\tau\right).
\end{equation}
This ensures  consistency of the derivation. Indeed, the path-multiplicative exponents are uniquely determined up to an integer shift, implying that the \textit{consistency condition} is independent of the particular Frobenius solution considered. Furthermore, this implies that the same expansion derived above is also valid for the other Frobenius solution, where the cosine function is simply replaced by the sine function. The solution with initial condition $\psi_e(0) = 0$ is then given by
\begin{equation}
    \psi^{\text{II}}_e(t) = \exp{\left[ i\frac{2f}{\omega}\sin^2\left(\frac{\omega t}2\right)\right]}H\ell^{(1/2)}_{\{0\}}\left(\boldsymbol{\eta};\sin^2\left(\frac{\omega t}2\right)\right).
\end{equation}

\section{Conclusions and outlook}
\label{sec:ceo}
We have established a direct analytical connection between the solutions of the linearly driven TLS and its RWA approximation, by mapping the corresponding Schrödinger problems to the CHE and the HE respectively. In particular, this connection is achieved by studying the local solutions in terms of their expansions in hypergeometric functions.
For the HE, this expansion takes the form of a single-sided series which, for the parameter values corresponding to the TLS with RWA driving, naturally truncates at first order, yielding the solutions known  for this system from standard treatments in quantum mechanics. In the case of the CHE, the corresponding expansion requires the introduction of PMF solutions and results in a bilateral series. This leads to the definition of two continued fractions, supplemented by an appropriate boundary condition, that we called \textit{consistency condition}, which determines the solution.
By expanding the continued fractions in the relevant parameter regimes and enforcing this consistency condition, we were able first to determine the path exponent and then to analytically recover the RWA approximation of the solutions. Considering the contributions up to first order, we recovered in particular the weak-driving and high-frequency limits correctly identifying the Stark and Bloch-Siegert corrections. The approach, however, allows to obtain from the exact solution expansions to an arbitrary perturbative order.

Our work recovers relevant limits of the linearly driven two-level system from exact analytical solutions, highlighting the role of path-multiplicative Floquet solutions to Heun equations, which has been largely overlooked in the literature. PMF solutions could also be employed to identify two-level systems with specific driving protocols that admit tractable analytical solutions. This can be achieved by considering parameters for which the hypergeometric series expansion of the local solutions truncates at finite order, thereby allowing for an exact analytical study of the system dynamics. Most importantly, the method here developed, based on perturbative expansions of continued fractions with appropriate boundary conditions, is of general validity and well-suited for applications in the diverse contexts where Heun equations naturally arise.

\appendix

\section{Coefficients in three-term recurrence relations}
\label{appendix:A}
In this Appendix we provide the expressions for the coefficients appearing in the three-term recurrence relations that must be solved in order to obtain the local solutions of the different Heun equations. These coefficients are directly expressed in terms of the parameters of the equations.
We first consider the recurrence relations for the HE, and later those for the CHE.

The general expression for $P_n^*, S_n^*,R_n^*$ in terms of the coefficients of the HE can be obtained following~\cite{Ronveaux1995} and reads
\begin{eqnarray}
    &&P_n^* = -\frac{(2\alpha +n-\gamma -\delta)(\alpha+\beta+n-\gamma-\delta)(\alpha+n-\delta)(\alpha+n-1)}{
    (2\alpha +2n -1 -\gamma-\delta)(2\alpha +2n -\gamma-\delta)
    } ,\nonumber \\
    &&S^*_n = \frac{Z_n}{(2\alpha +2n +2-\gamma-\delta)(2\alpha + 2n - \gamma - \delta)} -a(\alpha +n)(\alpha +n +1 - \gamma-\delta) -q ,\nonumber \\
    &&R^*_n = -\frac{(\alpha + n +2-\gamma-\delta)(n +1)(\alpha-\beta + n +1)(\alpha + n + 1-\gamma)}{(2\alpha + 2n +3 -\gamma - \delta)(2\alpha + 2n + 2-\gamma - \delta)},
\end{eqnarray}
with
\begin{eqnarray}
    Z_n &=&[(\alpha + n)(\alpha + n+1 - \gamma-\delta )+ \alpha\beta][2(\alpha +n)(\alpha + n +1 - \gamma - \delta)+\gamma(\gamma+ \delta -2)]+ \nonumber \\
    && +\epsilon(\alpha + n)(\alpha + n+1 -\gamma - \delta )(\gamma- \delta) .
\end{eqnarray}
Note that Eq.~(4.2.17c) on p.~50 of Ronveaux~\cite{Ronveaux1995}, where the notation $\omega = \gamma+ \delta -1$ was used, contains a typo: its denominator should read $(2\nu+\omega)(2\nu+\omega+1)$.

The general expressions for $A_n$, $B_n$, and $C_n$ in terms of the parameters of the CHE can be found in~\cite{Ronveaux1995} as well, and reads in our notation:
\begin{eqnarray}
A_n &=& -4p\frac{\left(n+\nu-2+\gamma+\delta\right)\left(n+\nu-1+\gamma\right)\left(n+\nu-1+\alpha\right)}{\left(2n +2\nu-3+\delta+\gamma\right)\left(2n + 2\nu-2+\gamma+\delta\right)},\nonumber \\
    B_n &=& \left(n+ \nu\right)\left(n + \nu+\gamma+\delta -1\right) -\sigma+ \nonumber \\
    &&\ +2p\left[\frac{\left(\gamma-\delta\right)\left[\alpha\left(\gamma+\delta-2\right)+2\left(n+\nu\right)\left(n+\nu+\gamma+\delta-1\right)\right]}{\left(2n +2\nu+\gamma+\delta-1\right)^2-1}+\alpha\right], \nonumber \\
    C_n &=& 4p\frac{\left(n+\nu+1\right)\left(n+\nu+\delta\right)\left(n+\nu+\delta+\gamma-\alpha\right)}{\left(2n +2\nu+\delta+\gamma\right)\left(2n + 2\nu+1+\gamma+\delta\right)}.
\end{eqnarray}
Note further that in Eq.~(2.3.33) on p.~104 of  Ronveaux~\cite{Ronveaux1995} the coefficient $b_{n,n}$ should be replaced by $2b_{n,n}$.






\bibliographystyle{apsrev4-2}
\bibliography{biblio}

\end{document}